\documentclass[useAMS,usenatbib]{mn2e}
\input{epsf}
\usepackage{amssymb}
\usepackage{natbib}
\usepackage{color}
\usepackage{journals}

\newbox\grsign \setbox\grsign=\hbox{$>$} \newdimen\grdimen \grdimen=\ht\grsign
\newbox\simlessbox \newbox\simgreatbox
\setbox\simgreatbox=\hbox{\raise.5ex\hbox{$>$}\llap
     {\lower.5ex\hbox{$\sim$}}}\ht1=\grdimen\dp1=0pt
\setbox\simlessbox=\hbox{\raise.5ex\hbox{$<$}\llap 
     {\lower.5ex\hbox{$\sim$}}}\ht2=\grdimen\dp2=0pt

\newcommand{\hMpc}{{\ifmmode{h^{-1}{\rm Mpc}}\else{$h^{-1}$Mpc }\fi}}
\newcommand{\hGpc}{{\ifmmode{h^{-1}{\rm Gpc}}\else{$h^{-1}$Gpc }\fi}}
\newcommand{\hkpc}{{\ifmmode{h^{-1}{\rm kpc}}\else{$h^{-1}$kpc }\fi}}
\newcommand{\hMsun}{{\ifmmode{h^{-1}{\rm {M_{\odot}}}}\else{$h^{-1}{\rm{M_{\odot}}}$}\fi}}
\newcommand{\Msun}{{\ifmmode{{\rm {M_{\odot}}}}\else{${\rm{M_{\odot}}}$}\fi}}

\title[Mapping between redshift and scale factor]{Testing the mapping between redshift and cosmic scale factor}
\author[R. Wojtak \& F. Prada]{Rados{\l}aw Wojtak$^{1, 2}$\thanks{E-mail: wojtak@stanford.edu, f.prada@csic.es} and Francisco Prada$^{1,3,4,5}$ \\
$^1$Kavli Institute for Particle Astrophysics and Cosmology, Stanford University, SLAC National Accelerator Laboratory, \\ Menlo Park, CA 94025, USA \\
$^2$ Dark Cosmology Centre, Niels Bohr Institute, University of Copenhagen, 
2100 Copenhagen \O, Denmark \\
$^3$ Instituto de F{\'i}sica Te{\'o}rica, (UAM/CSIC), Universidad Aut{\'o}noma de Madrid, Cantoblanco, E-28049 Madrid, Spain \\
$^4$ Campus of International Excellence UAM+CSIC, Cantoblanco, E-28049 Madrid, Spain \\
$^5$ Instituto de Astrof{\'i}sica de Andaluc{\'i}a, (IAA/CSIC), Glorieta de la Astrono{\'i}a, E-18190 Granada, Spain
\
}
\begin{document}

\maketitle

\begin{abstract}
The canonical redshift$-$scale factor relation, $1/a=1+z$, is a key element in the standard $\Lambda$CDM model of the big bang cosmology. 
Despite its fundamental role, this relation has not yet undergone any observational tests since Lema$\hat{\imath}$tre and Hubble established the expansion of the Universe. 
It is strictly based on the assumption of the Friedmann--Lema$\hat{\imath}$tre--Robertson--Walker metric describing a locally homogeneous and isotropic universe and that photons 
move on null geodesics of the metric. Thus any violation of this assumption, within general relativity or modified gravity, can yield a different mapping between 
the model redshift $z=1/a-1$ and the actually observed redshift $z_{\rm obs}$, i.e. $z_{\rm obs}\neq z$. Here we perform a simple test of consistency for the standard redshift$-$scale factor relation by determining simultaneous observational constraints on the concordance $\Lambda$CDM cosmological parameters and a generalized redshift mapping $z=f(z_{\rm obs})$. Using current baryon acoustic oscillations (BAO) and Type Ia supernova (SN) data we demonstrate that the generalized redshift mapping is strongly degenerated with dark energy. Marginalization over a class of monotonic functions $f(z_{\rm obs})$ changes 
substantially degeneracy between matter and dark energy density: the density parameters become anticorrelated with nearly vertical axis of degeneracy. Furthermore, we show that 
current SN and BAO data, analysed in a framework with the generalized redshift mapping, do not constrain dark energy unless the BAO data include the measurements from the Ly$\alpha$ forest of high-redshift quasars.

\end{abstract}


\section{Introduction}

A cornerstone of the big bang model of cosmology is the phenomenon of redshifted light. There is no doubt that observed redshifts of distant galaxies reflect expansion of the Universe \citep[][]{Lem1927,Hub1929}. However, it is less obvious how the observed redshifts relate to the successive phases of the expansion. The current standard cosmological model describes the evolution of the Universe in terms of homogeneous space expansion, which is quantified by a single function of time, the so-called cosmic scale factor $a(t)$. In this theoretical framework, the observed redshifts $z_{\rm obs}$ of distant galaxies map on to the scale factor through the standard formula $1+z_{\rm obs}=1/a$. Despite the fact that it underlies all cosmological fits and conversions between evolution of cosmological observables parametrized by the scale factor and the same evolution parametrized 
by cosmological redshift, this relation does not have a status of a fundamental theory. It depends solely on the assumption that propagation of light in the Universe can be accurately described within theoretical framework of general relativity (GR) with the Friedmann--Lema$\hat{\imath}$tre--Robertson--Walker (FLRW) metric. Mapping between observed redshifts and the evolution of the Universe may take a different form if the Universe violates this assumption. Deviation from the conventional redshift--scale factor mapping may emerge in some theories of modified gravity, for example $\tilde{\delta}$ gravity \citep[see e.g.][]{Alf2012,Alf2013}, or simply from inadequacy of the FLRW metric to fit our true inhomogeneous Universe \citep[][]{Ell1987,Cli2009,Cli2012,Lav2013}. A very simple example of the latter possibility, although giving rise to only a sub-percent effect in measurements of cosmological parameters, is the local gravitational redshift \citep{Woj2015}. These theoretical arguments fully motivate and justify the notion of observational tests aimed to corroborate 
the canonical relation between observed redshifts and cosmic scale factor. Unlike all standard approaches in observational cosmology, this kind of test is not supposed to assume any theoretically motivated redshift$-$scale factor mapping, but rather to determine it empirically.

The most puzzling problem of modern cosmology is arguably the nature of dark energy, a mysterious component of the Universe responsible for its apparent acceleration; and inferred from observations of Type Ia supernovae \citep[SN; see][]{Rie1998, Per1999}. An undeniable success of the concordance cosmological $\Lambda$ cold dark matter ($\Lambda$CDM) model is the fact that the same form of dark energy manifests itself consistently in various cosmological measurements, from the cosmic microwave background radiation \citep[CMB; see e.g.][]{Pla2014} and baryon acoustic oscillations \citep[BAO; see e.g.][]{And2012,Bla2011} to the abundance of galaxy clusters \citep[see e.g.][]{Man2008,Vik2009}. Yet, this solid empirical background, which assumes Einstein's GR as gravity theory on cosmological scales, is unfortunately not accompanied by any advance in our theoretical understanding of this phenomenon. 
Lack of compelling theoretical explanations on the one hand and vast observational data on the other motivate many researchers for pursuing challenging observational tests of all relevant ingredients of the standard $\Lambda$CDM model which constitute a framework for our reasoning on the matter-energy content and expansion history of the Universe. 

Current and ongoing tests aim to verify, among the fundamental pillars of the standard cosmological model, GR \citep[see e.g.][]{Rap2009,Cat2015}, homogeneity of the matter density distribution on large scales \citep[see e.g.][]{Scr2012} or the distance duality relation used to convert between angular diameter and luminosity distances \citep[see e.g.][]{Hol2010,Ell2013}. Having mentioned this, it seems quite surprising to notice that, since Lema$\hat{\imath}$tre and Hubble established the expansion of the Universe almost a century ago, the mapping between observed redshifts and cosmic scale factor have not undergone stringent observational tests yet and it is commonly taken for granted.

The first attempt to test the redshift$-$scale factor relation was pursued by \citet{Bas2013}. However, as we shall show in section 2, the theoretical framework built for this test turns out to be inconsistent with the distance duality relation which has a far more fundamental status than mapping between redshifts and cosmic scale factor \citep[see e.g.][]{Ellis2007}. In this paper we develop a new approach to testing the redshift$-$scale factor mapping in a way which obeys the distance duality relation. We propose to use best available observations of Type Ia SN and BAO (angular diameter distances and Hubble parameters) to place constraints on the mapping and thus to test its commonly used canonical form, i.e. $1+z_{\rm obs}=1/a$. Our proposition goes beyond classical tests aimed at discriminating between metric and nonmetric origin of cosmological redshifts. The latter, such as Zwicky's hypothesis of `tired light', was ruled out at high confidence level by observations of the aging rate of Type Ia SN \citep{Blo2008}. In our test, we explicitly employ metric-based interpretation of the observed redshifts and the question we address is whether the standard mapping between cosmological redshifts and cosmic scale factor is consistent with observations. It is also worth noting that our approach retains pristine relation between the CMB temperature and observed redshift. This relation has been extensively tested using measurements of 
the Sunyaev$-$Zeldovich effect \citep{Sar2014,Mar2015,Luz2015}.

The manuscript is organized as follows. In section 2, we describe a new phenomenological model generalizing the standard redshift$-$scale factor mapping. This model is then used to place observational constraints on the mapping based on SN and BAO data. Description of the data as well as data analysis technical details are outlined in section 3. Results are presented in section 4, which is then followed by discussion in section 5. Summary is given in section 6.

\section{Model}
Without loss of generality we assume that all measured redshifts are corrected for peculiar velocities of the observed galaxies and our own reference frame. This can be easily achieved by applying smoothing or binning in redshift space so that peculiar velocities are effectively averaged out. For the sake of a clear presentation of the methodology and modelling, 
we distinguish the actually \textit{observed redshift}
\begin{equation}
z_{\rm obs}=\frac{\lambda_{\rm obs}-\lambda_{\rm em}}{\lambda_{\rm em}},
\label{z_zobs}
\end{equation}
where $\lambda_{\rm em}$ and $\lambda_{\rm obs}$ are wavelengths of photons at the point of emission and observation, 
from the \textit{model redshift} $z$ due to homogenous expansion of space. The former, given in eq.~(\ref{z_zobs}), 
is an observable with the exact relativistic definition given by
\begin{equation}
z_{\rm obs}=\frac{(u^{\alpha}k_{\alpha})_{\rm em}}{(u^{\alpha}k_{\alpha})_{\rm obs}}-1,
\label{z_obs_gen}
\end{equation}
where $u^{\alpha}$ is the four-velocity vector of the observer and $k^{\alpha}$ is the tangent vector to the null geodesic line of the photon. The latter, given by
\begin{equation}
z=\frac{1}{a}-1,
\label{z_a}
\end{equation}
where $a$ is cosmic scale factor in the FLRW metric describing a homogenous and isotropic expanding universe, is a special case of eq.~(\ref{z_obs_gen}). 

The concordance $\Lambda$CDM cosmological model does not distinguish between $z_{\rm obs}$ and $z$. Hereafter, we shall refer to this assumption as the standard redshift mapping, i.e. $z_{\rm obs}=z$. The standard redshift mapping assumes implicitly general framework of relativity (photons propagating on null geodesics) and the FLRW metric. 
The former assumption leads to eq.~(\ref{z_obs_gen}) and the latter implies eq.~(\ref{z_a}). Violation of any 
of these two assumptions can modify the standard relation between $z$ and $z_{\rm obs}$. Examples of such violations comprise some modifications of GR, e.g. $\tilde{\delta}$ gravity \citep{Alf2013}, or inhomogeneous solutions of Einstein's equations \citep[see e.g.][]{Cli2009,Cli2012}. Hereafter, we shall refer to any modification of $z_{\rm obs}=z$ relation as redshift remapping. The remapping is a monotonic function $z=f(z_{\rm obs})$ and the standard redshift mapping is its special case when $f$ is an identity function.

The theoretical framework of our model is built upon the assumption that the observed redshift $z_{\rm obs}$ is fully determined by the true metric of the Universe. We therefore 
exclude all possible models based on a nonmetric origin of cosmological redshift such as Zwicky's hypothesis of `tired light'. Metric-based interpretation of observed redshifts guarantees agreements with three fundamentals observational facts: blackbody spectrum of the CMB \citep{Fix1996}, redshift evolution of the CMB temperature 
\citep{Sar2014,Mar2015,Luz2015} and redshift dependence of the SN aging rate \citep{Blo2008}.

\begin{figure*}
\begin{center}
    \leavevmode
    \epsfxsize=16cm
    \epsfbox[150 65 1290 580]{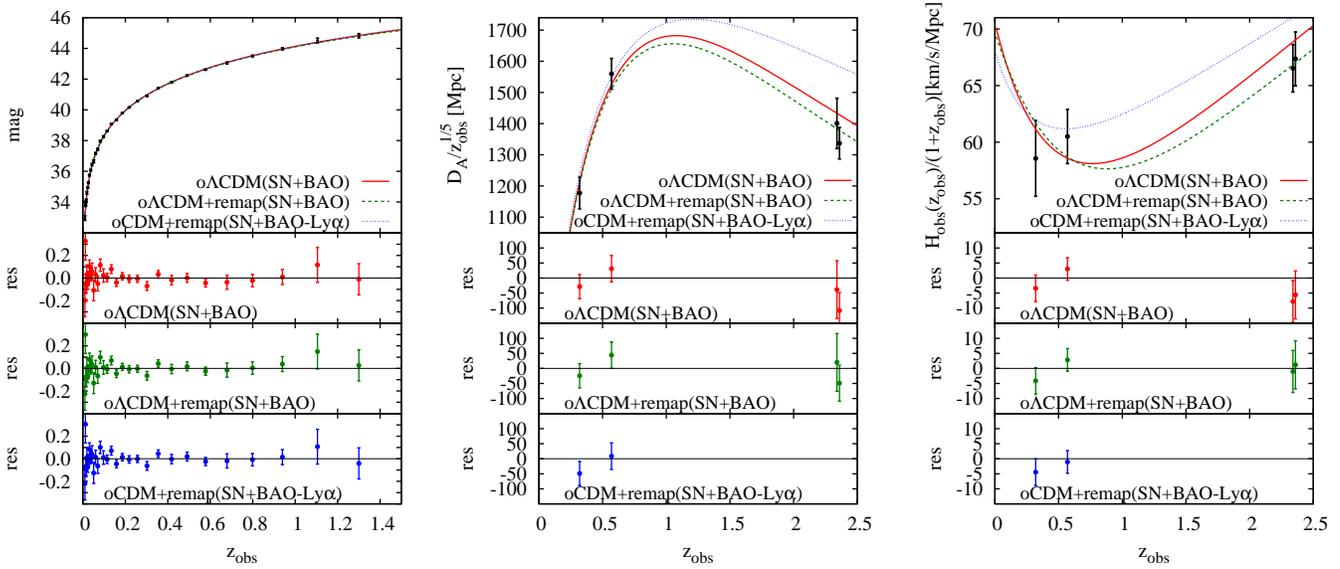}
\end{center}
\caption{Apparent magnitudes of Type Ia supernovae (SN) and measurements of baryon acoustic oscillations (BAO): angular diameter distances $D_{\rm A}(z_{\rm obs})$ and 
Hubble parameters $H_{\rm obs}(z_{\rm obs}$). The measurements are compared to best-fit models: open $\Lambda$CDM (solid red lines) with the standard redshift mapping ($z_{\rm obs}=z$), open $\Lambda$CDM (dashed green lines) and open CDM $-$ no dark energy (dotted blue lines), both with redshift remapping given by eq.~(\ref{remapping}). Both $\Lambda$CDM models are fitted to the best available entire SN and BAO data sets, while fitting the CDM model omits the Ly$\alpha$ forest BAO measurements of SDSS-III/BOSS quasars at $z=2.34,\;2.36$.
}
\label{residuals}
\end{figure*}

\subsection{Redshift remapping}
A simple observational test of the standard redshift mapping can be devised by considering a cosmological model with a more general form of the relation between $z$ and $z_{\rm obs}$. Here we propose the following phenomenological relation between $z$ and $z_{\rm obs}$:
\begin {equation}
\frac{z}{z_{\rm obs}}=1+\frac{\alpha}{1+z_{\rm obs}},
\label{remapping}
\end{equation}
where $\alpha$ is a free parameter. The proposed parametrization may not exploit all possibilities; however, it has a number of well justified properties. First of all, deviation from the standard redshift mapping is governed by a single parameter $\alpha$ (with the standard redshift mapping $z_{\rm obs}=z$ recovered for $\alpha=0$), which prevents an unnecessary 
expansion of the explored parameter space. In this regard, the proposed model of redshift remapping is a one-parameter extension of a standard $\Lambda$CDM cosmological model. 
Secondly, the model does not introduce strong limitations on a relation between $z$ and $z_{\rm obs}$ and allows for both $z<z_{\rm obs}$ and  $z>z_{\rm obs}$. Finally, this model recovers the standard redshift mapping at high redshifts, i.e. $z/z_{\rm obs}\rightarrow 1$ for $z_{\rm obs}\rightarrow\infty$. This property ensures that the new redshift remapping is unlikely to affect the interpretation of the CMB observations. This is a very conservative 
assumption, although not strictly necessary. We note, however, that fitting any kind of redshift remapping may in general change the Universe's expansion history and thus the anisotropies of the CMB, as we discuss in Sec. 5.

The calculation of all basic observables such as distances and the Hubble parameter $H$ requires using the equations that govern the expansion of space. We assume that in a $\Lambda$CDM cosmology with dark energy equation of state $w=-1$, the expansion rate of the Universe on large scales is adequately described by the Friedmann equation,  
\begin{equation}
H^{2}(z)/H_{0}^{2}=\Omega_{\rm m}(1+z)^{3}+\Omega_{\rm k}(1+z)^{2}+\Omega_{\Lambda},
\label{Friedmann}
\end{equation}
where $1+z=1/a$. $\Omega_{\rm m}$, $\Omega_{\rm k}$ and $\Omega_{\Lambda}$ are the density parameters for the matter (cold dark matter and baryons), curvature and the cosmological constant. The radiation term has been neglected since observations reported very small radiation density $\Omega_{\rm rad}$ in the redshift range of current SN and BAO observations.

We note that invoking the Friedmann equation in our context may be logically inconsistent with some possible reasons of violating the standard redshift mapping. On the other hand, this approach is independent of any specific modification of the Friedmann equation emerging from possible new theoretical frameworks capable of predicting a non-standard redshift remapping. Overall, we think that combining the Friedmann equation with redshift remapping is general enough to think of the proposed analysis as a test of consistency between observations and the standard theoretical framework for modelling the expansion history of the Universe as well as establishing the mapping between observed redshifts and cosmic scale factor at the same time. We also note that an alternative approach to testing redshift remapping without the assumption of GR and the Friedmann equation is presented in \citet[][]{Bas2013}.

\subsection{Cosmological distances}

The comoving distance to an object observed at redshift $z_{\rm obs}$ is given by
\begin{equation}
d_{\rm C}(z_{\rm obs})=c\int_{0}^{f(z_{\rm obs})}\frac{\textrm{d}x}{H(x)},
\label{dC}
\end{equation}
where $z=f(z_{\rm obs})$ is redshift remapping given by eq.~(\ref{remapping}) and $H(x)$ is the Hubble parameter given by the Friedmann equation (\ref{Friedmann}). It is apparent that the standard formula for $d_{\rm C}$ \citep[see e.g.][]{Ell1971, Hogg1999}can be recovered by using $f(z_{\rm obs})=z_{\rm obs}$. Once the comoving distance is determined, the luminosity distance $d_{\rm L}$, and the angular diameter distance $d_{\rm A}$ are calculated as follows,
\begin{eqnarray}
d_{\rm L}(z_{\rm obs})  & = &  (1+z_{\rm obs}) \, d_{\rm H} \, f_{\rm C}[d_{\rm C}(z_{\rm obs})/d_{\rm H} ]\nonumber \\
d_{\rm A}(z_{\rm obs})  & = &  \frac{1}{1+z_{\rm obs}} \, d_{\rm H} \, f_{\rm C}[d_{\rm C}(z_{\rm obs})/d_{\rm H}],
\label{dLdA}
\end{eqnarray}
where $d_{\rm H}=(c/H_{0})/\sqrt{|\Omega_{k}|}$ and
\begin{equation}\label{f(x)}
	f_{\rm C}(x)=  \left\{
	\begin{array}{lll}
	 \sinh(x) & \Omega_{\rm k}>0,\\
	 x & \Omega_{\rm k}=0,\\
	 \sin(x) & \Omega_{\rm k}<0.\\
\end{array} \right.
\end{equation}
The above equations resemble closely the standard formulae for cosmological distances described in e.g. \citet{Hogg1999}. The only non-trivial part of them are the $(1+z_{\rm obs})$ 
and $1/(1+z_{\rm obs})$ factors. It may not be so obvious why the redshift variable used in these factors is not $z$ given by eq.~(\ref{z_a}). This becomes clear when we realize that using the model redshift $z$ instead of the actually observed redshift $z_{\rm obs}$ would violate the so-called distance duality relation, i.e.

\begin{equation}
d_{\rm L}(z_{\rm obs})=(1+z_{\rm obs})^{2}d_{\rm A}(z_{\rm obs}).
\label{duality}
\end{equation}

This relation is a direct consequence of photon number conservation and the reciprocity theorem \citep{Eth1933}. It holds for general metric theories of gravity in which light travels along null geodesics in Lorentzian spacetime. Redshift $z_{\rm obs}$ in this relation is the actually observed/measured redshift of photon sources \citep[see e.g.][]{Ellis2007,Bas2004}. It is worth noting that the cosmic distance duality relation holds also in $\tilde{\delta}$ gravity, which is not a metric theory \citep{Alf2012}.

Our redshift remapping does not change the distance-duality relation between the luminosity distance and the angular diameter distance given in eq.~(\ref{duality}). This ensures that observational constraints on the remapping from luminosity distance measures, as provided by Type Ia supernova data, are expected to coincide with those from BAO angular diameter distance measurements unless there is an external source of tension between these two data sets. This self-consistency of the model fails when one attempts to violate the distance-duality relation by using the $(1+z)$ and $1/(1+z)$ factors in eq.~(\ref{dLdA}) instead of $(1+z_{\rm obs})$ and $1/(1+z_{\rm obs})$. This approach was recently adopted by \citet{Bas2013}. As shown in their analysis of different compilations of cosmological data sets, observational constraints on redshift remapping from SN luminosity distances cannot be reconciled with analogous constraints from BAO angular diameter distances.

\subsection{Hubble parameter}

Observational determinations of the Hubble parameter $H_{\rm obs}$ from BAO rely on measuring a redshift range  $\textrm{d}z_{\rm obs}$ corresponding to the comoving BAO scale $\textrm{d}r$. The Hubble parameter is then estimated as a ratio $\textrm{d}z_{\rm obs}/\textrm{d}r$. For the standard redshift mapping ($z_{\rm obs}=z$), the observed Hubble 
parameter $H_{\rm obs}(z_{\rm obs})$ can be directly modelled by the Hubble parameter given by the Friedmann equation, i.e. $H_{\rm obs}(z_{\rm obs})=H(z)$. Redshift remapping, however, changes the relation between these two Hubble parameters in the following way:
\begin{equation}
H_{\rm obs}(z_{\rm obs})=H[f(z_{\rm obs})]/\Big(\frac{\textrm{d}f(z_{\rm obs})}{\textrm{d}z_{\rm obs}}\Big). 
\end{equation}
The observed Hubble parameter $H_{\rm obs}(z_{\rm obs})$ is therefore never equal to the Hubble parameter given by the Friedmann equation at the corresponding redshift $z=f(z_{\rm obs})$ unless derivative $\textrm{d}f/\textrm{d}z_{\rm obs}$ equals to $1$. For the 
redshift remapping adopted in our work, the Hubble parameter measured in observations is given by
\begin{equation}
H_{\rm obs}(z_{\rm obs})=\frac{H[f(z_{\rm obs})]}{1+\alpha/(1+z_{\rm obs})^{2}}.
\end{equation}
This implies that the observed Hubble constant $H_{\rm obs}(0)$ scales with the global Hubble constant defined in the Friedmann equation as $H_{\rm obs}(0)=H_{0}/(1+\alpha)$. Therefore, a non-zero value of parameter $\alpha$ in our redshift remapping model gives rise to a difference between the locally measured Hubble constant and its counterpart 
defined in the Friedmann equation. However, that this is not a generic feature of all possible redshift remappings among which one can find examples with $H_{\rm obs}(0)=H_{0}$.

\section{Observational constraints}

\begin{table*}
\begin{center}
\begin{tabular}{cccc}
$z_{\rm obs}$ & $H_{\rm obs}r_{\rm d}\;(10^{3}\textrm{km}\;\textrm{s}^{-1})$ & $d_{\rm A}/r_{\rm d}$ & Reference \\
\hline
$0.32$ & $11.33\pm0.56$ & $6.33\pm0.19$ & \citet{Gil2015} \\
$0.32$ & $11.44\pm0.74$ & $6.40\pm0.36$ & \citet{Cue2015} \\
$0.57$ & $13.84\pm0.43$ & $9.42\pm0.15$ & \citet{Gil2015} \\
$0.57$ & $14.14\pm0.68$ & $9.51\pm0.45$ & \citet{Cue2015} \\
$2.34$ & $32.72\pm1.03$ & $11.28\pm0.65$ & \citet{Del2015} \\
$2.36$ & $33.33\pm1.18$ & $10.78\pm0.41$ & \citet{Fon2014} \\
\end{tabular}
\caption{SDSS-III/BOSS DR12 measurements of the BAO angular distances and Hubble parameters used in this work. $r_{\rm d}$ is the sound horizon scale at the drag epoch.}
\label{baodata}
\end{center}
\end{table*}

Our redshift remapping model is a one-parameter extension of the standard $\Lambda$CDM cosmological model. Here we use a compilation of data sets comprising observations of Type Ia SN and anisotropic BAO measurements of angular distances and Hubble parameters to place constraints on the redshift remapping and cosmological parameters, and to assess possible improvement of a cosmological fit based on these data sets. Apparent magnitudes of SN come from a joint analysis of SDSS-II and SNLS supernova samples carried out by \citet{Bet2014}. The data set comprises mean apparent magnitudes of Type Ia SN measured in $31$ redshift bins (see the top left panel of Fig.~\ref{residuals}) and the covariance matrix. The Hubble parameters and the angular diameter distances measured from BAO observations are summarized in Table~\ref{baodata}. We use the most recent measurements based on the anisotropic galaxy clustering analysis performed on the final Data Release 12 of the SDSS-III Baryon Oscillation Spectroscopic Survey (BOSS) LOWZ ($z_{\rm obs}=0.32$) and CMASS ($z_{\rm obs}=0.57$) Luminous Red Galaxy samples. In our cosmological fits, we employ averages of two galaxy clustering measurements for each LOWZ and CMASS samples based on two independent techniques, an analysis of the anisotropic fitting of the two-point correlation function \citep{Cue2015} 
and the power spectrum in Fourier space \citep{Gil2015}. Finally, we also make use of the BAO angular diameter distance and Hubble parameter measured from the flux-correlation of the Ly$\alpha$ forest of SDSS-III/BOSS high-redshift quasars \citep{Del2015} and from the cross-correlation of quasars with the Ly$\alpha$ forest absorption \citep{Fon2014}. Both measurements were made using a sample of quasars from BOSS Data Release 11. 
In order to convert dimensionless estimates of the BAO signal into a physical scale, we adopt 
a fiducial model with the sound horizon scale at the drag epoch $r_{\rm d\; fid}=147.27\;{\rm Mpc}$, consistent with CMB Planck observations \citep{Pla2015}. The resulting values of the angular diameter distances and Hubble parameters are plotted in the top panels of Fig.~\ref{residuals}. As we explain below, using a fixed value of $r_{\rm d}$ does not make the results of our analysis dependent on Planck cosmology. Since we treat normalization of the BAO angular diameter distance and Hubble parameter as a nuisance parameter, 
our results are essentially independent of cosmological constraints from the CMB.

Our cosmological model contains three groups of parameters: cosmological parameters (density parameters $\Omega_{\rm m}$ and $\Omega_{\Lambda}$), 
redshift remapping parameter $\alpha$ and two nuisance parameters, $M_{\rm SN}$ and $M_{\rm BAO}$, which are used to normalize the magnitude$-$redshift 
relation for Type Ia SN data,
\begin{equation}
m(z_{\rm obs})=5\log10[d_{\rm L}(z_{\rm obs})H_{0}]+M_{\rm SN},
\end{equation}
and the BAO data sets (both distances and Hubble parameters). We define the BAO normalization parameter in the following way:
\begin{equation}
M_{\rm BAO}=h_{70}\frac{r_{\rm d}}{r_{\rm d\;fid}},
\end{equation}
where the Hubble constant $H_{0}=h_{70}70{\rm\;km\;s^{-1}\;Mpc^{-1}}$ and $r_{\rm d\;fid}=147.27{\rm\;Mpc}$. The Hubble constant is included in both normalization parameters and therefore it cannot be constrained directly from the data unless prior knowledge of the absolute magnitude of the supernovae or the sound horizon scale is assumed. We take maximally 
conservative approach and adopt flat priors on the two normalizations, $M_{\rm SN}$ and $M_{\rm BAO}$, as well as the remaining parameter of the model. All final constraints are marginalized over the normalization parameters and therefore they are independent of the adopted fiducial value of the sound horizon scale.

We find the best-fitting parameters by maximizing the likelihood function $L$ given by
\begin{equation}
\ln L(\theta|{\rm data}) \propto -\frac{\chi^{2}_{\rm SN}}{2}-\frac{\chi^{2}_{\rm BAO}}{2},
\label{likelihood}
\end{equation}
and
\begin{eqnarray}
\chi^2_{\rm SN} & = & \sum_{i,j}^{31}C^{-1}_{ij}[m_{i}-m(z_{{\rm obs}\;i},\theta)][m_{j}-m(z_{{\rm obs}\;j},\theta)]\nonumber, \\
\chi^{2}_{\rm BAO} & = & \sum_{i}^{4}\frac{[d_{{\rm A}\;i}-d_{\rm A}(z_{\rm obs\; i},\theta)]^{2}}{\sigma_{{\rm A}\;i}^{2}}\nonumber \\
 & + & \sum_{i}^{4}\frac{[H_{{\rm obs}\;i}-H_{\rm obs}(z_{\rm obs\; i},\theta)]^{2}}{\sigma_{{\rm H}\;i}^{2}},
\label{chi2}
\end{eqnarray}
where $\theta$ is a vector of the model parameters, $C$ is the covariance matrix for the SN data set, $\sigma_{\rm A}$ and $\sigma_{\rm H}$ are the errors of the angular diameter 
distances and the Hubble parameters measured from BAO. We employ a Monte Carlo Markov Chain technique to explore the parameter space and thus to determine confidence regions of the model parameters. Markov chains are computed using the Metropolis$-$Hastings algorithm and the confidence regions (intervals) presented in all figures and tables are the $1\sigma$ and $2\sigma$ contours or intervals of the marginalized likelihood function.

In order to assess the potential improvement of the cosmological fit with both SN and BAO data sets, achieved by using our newly adopted redshift remapping, we compare the best-fitting 
models in terms of two criteria for model selection: the Bayesian information criterion (BIC) and the Akaike information criterion \citep[AIC; ][]{Tro2008}. Neglecting an additive constant given by the normalization of the likelihood functions, we evaluate the two criteria in the following way: ${\rm BIC}=(\chi^{2}_{\rm SN}+\chi^{2}_{\rm BAO})_{\rm min}+k\ln(n)$ 
and ${\rm AIC}=(\chi^{2}_{\rm SN}+\chi^{2}_{\rm BAO})_{\rm min}+2k$, where $k$ is the number of estimated parameters in the model, $n$ is the number of data points and $\chi^{2}_{\rm min}$ is the minimum value of $\chi^{2}$. The statistically more favoured model is the one which minimizes BIC or AIC.

\section{Results}

\begin{figure*}
\begin{center}
    \leavevmode
    \epsfxsize=16cm
    \epsfbox[99 70 858 410]{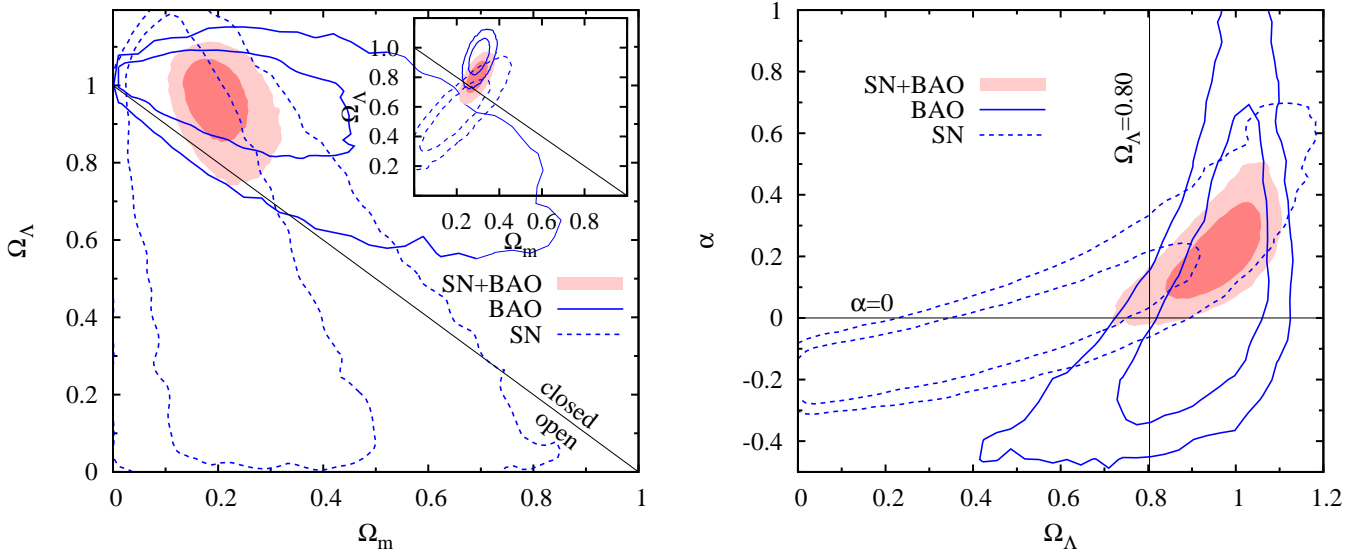}
\end{center}
\caption{Constraints on open $\Lambda$CDM model parameters with redshift remapping given by eq.~(\ref{remapping}). The solid and dashed blue lines show results from 
the best available BAO (all data points) and SN data, and the shaded contours results from an analysis combining both 
data sets ($1\sigma$ and $2\sigma$ contours). The inset panel shows constraints on the density parameters for an open $\Lambda$CDM model with 
the standard redshift mapping ($z_{\rm obs}=z$). The black vertical line on the right-hand panels indicates the best-fitting $\Omega_{\Lambda}$ for an open $\Lambda$CDM model with the standard redshift mapping. The new redshift remapping rotates axes of degeneracies on 
the $\Omega_{\rm m}-\Omega_{\Lambda}$ plane towards a line of flat models (black diagonal line). The redshift remapping is degenerated with the dark energy density parameter $\Omega_{\Lambda}$.}
\label{openLCDM}
\end{figure*}

\begin{table*}
\begin{center}
\begin{tabular}{cccccccc}
Model & $\Omega_{\rm m}$ & $\Omega_{\Lambda}$ & $\alpha$ & $\chi^{2}/{\rm dof}$ & BIC & AIC & $M_{\rm BAO}$ \\
\hline
o$\Lambda$CDM+$z=z_{\rm obs}$(SN+BAO) &  $0.292_{-0.032}^{+0.035}$ & $0.803_{-0.075}^{+0.069}$ & $\alpha=0$ & 1.24 & 58.05 & 51.40 & $1.008_{-0.025}^{+0.023}$\\
 & & & & & & & \\
o$\Lambda$CDM+$z\neq z_{\rm obs}$(SN+BAO) &  $0.189_{-0.028}^{+0.055}$ & $0.969_{-0.089}^{+0.055}$ & $0.225_{-0.120}^{+0.083}$ & 1.10 & 55.77 & 47.46 &$1.219_{-0.115}^{+0.077}$  \\
\hline
flat $\Lambda$CDM+$z=z_{\rm obs}$(SN+BAO) & $0.267_{-0.019}^{+0.020}$ & $\Omega_{\Lambda}=1-\Omega_{\rm m}$ & $\alpha=0$ & 1.23 & 55.39 & 50.40 & $1.006_{-0.023}^{+0.025}$\\
 & & & & & & & \\
flat $\Lambda$CDM+$z\neq z_{\rm obs}$(SN+BAO) & $0.192_{-0.035}^{+0.058}$ & $\Omega_{\Lambda}=1-\Omega_{\rm m}$ & $0.121_{-0.093}^{+0.079}$ & 1.21 & 56.84 & 50.18 & $1.120_{-0.091}^{+0.074}$ \\
\hline
oCDM+$z=z_{\rm obs}$(SN+BAO) & $0.079_{-0.028}^{+0.031}$ & $\Omega_{\Lambda}=0$ & $\alpha=0$ & 3.38 &132.8 & 127.9& $0.934_{-0.020}^{+0.021}$ \\
 & & & & & & & \\
oCDM+$z\neq z_{\rm obs}$(SN+BAO) & $0.478_{-0.092}^{+0.105}$ & $\Omega_{\Lambda}=0$ & $-0.250_{-0.030}^{+0.029}$ &1.95 & 82.94 & 76.29 &$0.746_{-0.030}^{+0.030}$\\
\end{tabular}
\caption{Constraints on the parameters of cosmological models with the standard 
redshift mapping ($z=z_{\rm obs}$) or with redshift remapping given by eq.~(\ref{remapping}) ($z\neq z_{\rm obs}$) and 
parametrized by $\alpha$, based on a joint analysis of the SN and BAO data sets. The table contains best-fitting values and $1\sigma$ confidence 
intervals of the parameters (density parameters $\Omega_{\rm m}$ and $\Omega_{\Lambda}$, parameter $\alpha$ of the redshift remapping and 
the normalization of the BAO signal, $M_{\rm BAO}$), reduced $\chi^{2}$ ($\chi^{2}/{\rm dof}$), the Bayesian information criterion (BIC) and the Akaike information criterion (AIC). 
The two top rows of the table present results for an open $\Lambda$CDM model (o$\Lambda$CDM, see also Fig.~\ref{openLCDM}), the middle two 
rows for a flat $\Lambda$CDM model (flat $\Lambda$CDM, $\Omega_{\rm m}+\Omega_{\Lambda}=1$) and the two bottom rows 
for an open CDM model without dark energy (oCDM, $\Omega_{\Lambda}=0$).
}
\label{fits}
\end{center}
\end{table*}

Fig.~\ref{openLCDM} shows constraints on the parameters of an open $\Lambda$CDM cosmological model with redshift remapping given by eq.~(\ref{remapping}). The solid and dashed contours show results from fitting independently the SN and BAO data 
by neglecting respectively $\chi^{2}_{\rm BAO}$ or $\chi^{2}_{\rm SN}$ in the likelihood function (\ref{likelihood}). The shaded contours demonstrate results from a joint analysis combining both data sets. The inset panel shows the corresponding constraints for a reference open $\Lambda$CDM model with the standard redshift mapping, i.e. $z_{\rm obs}=z$. Fig.~\ref{residuals} demonstrates how accurately the best-fitting models with the standard redshift mapping or with the new redshift remapping recover the measurements.

\begin{figure*}
\begin{center}
    \leavevmode
    \epsfxsize=16cm
    \epsfbox[99 70 858 410]{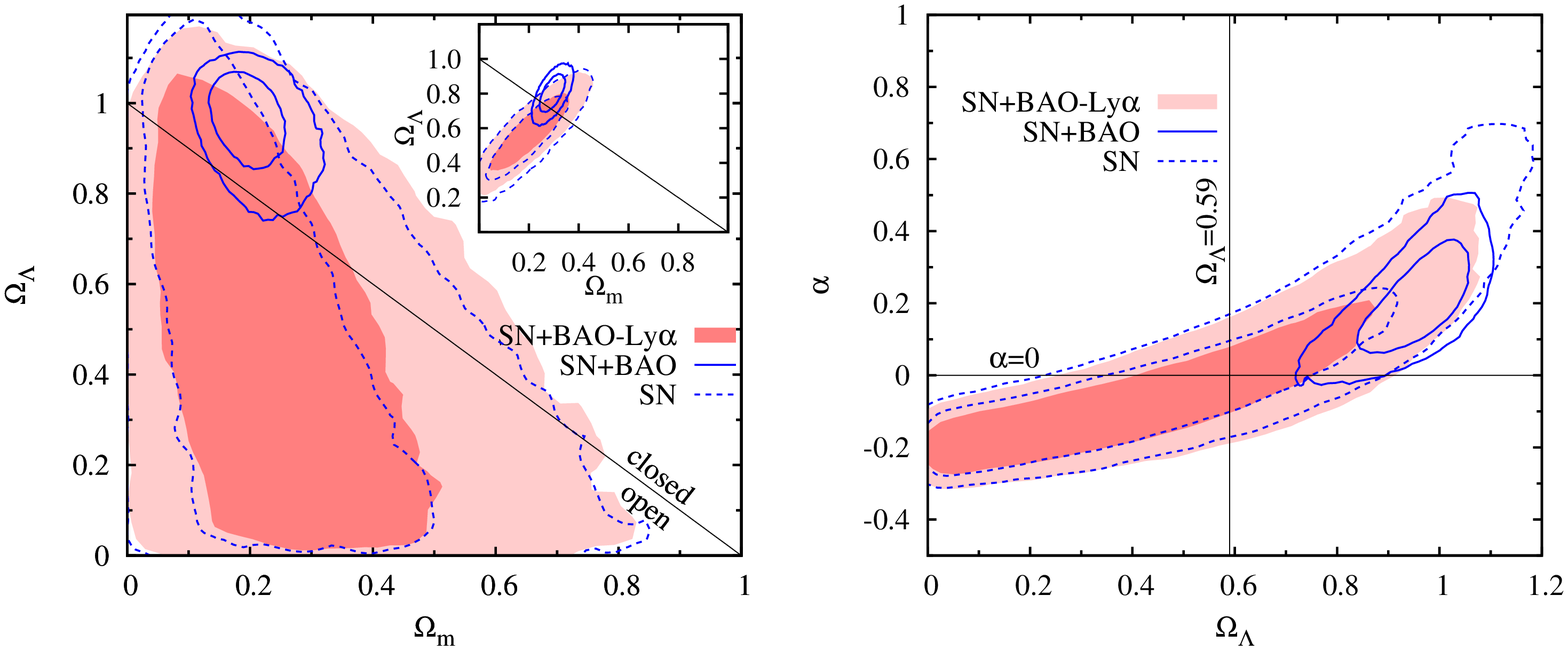}
\end{center}
\caption{Impact of BAO measurements from Ly$\alpha$ forest data of SDSS-III/BOSS quasars on the open $\Lambda$CDM model parameters with redshift remapping given by eq.~(\ref{remapping}). The solid blue lines and shaded contours show the results from an analysis combining the best available SN data set and BAO measurements, with or without considering the Ly$\alpha$ forest data points ($H_{\rm obs}$ and $D_{\rm A}$ at 
$z_{\rm obs}=2.34,\;2.36$), i.e. SN+BAO and SN+BAO$-$Ly$\alpha$. The dashed blue lines show constraints from the SN data alone. The inset panel shows constraints on the density parameters for an open $\Lambda$CDM model with the standard redshift mapping ($z_{\rm obs}=z$). The black vertical line on the right-hand panels indicates the best-fitting $\Omega_{\Lambda}$ for an open $\Lambda$CDM model with the standard redshift mapping. The omission of the BAO Ly${\alpha}$ data points in the analysis yields constraints which are fully consistent with a cosmological model $\it{without}$ dark energy and $\alpha<0$ (redshift remapping with $z<z_{\rm obs}$). The new axis of degeneracy between $\Omega_{\rm m}$ and $\Omega_{\Lambda}$ becomes nearly vertical.
}
\label{openLCDMa}
\end{figure*}

\begin{table*}
\begin{center}
\begin{tabular}{cccccccc}
Model & $\Omega_{\rm m}$ & $\Omega_{\Lambda}$ & $\alpha$ & $\chi^{2}/{\rm dof}$ & BIC & AIC & $M_{\rm BAO}$ \\
\hline
o$\Lambda$CDM+$z=z_{\rm obs}$(SN+BAO--Ly$\alpha$) &  $0.216_{-0.110}^{+0.096}$ & $0.593_{-0.159}^{+0.134}$ & $\alpha=0$ & 1.13 & 49.20 & 42.97 & $0.977_{-0.026}^{+0.026}$\\
 & & & & & & & \\
o$\Lambda$CDM+$z\neq z_{\rm obs}$(SN+BAO--Ly$\alpha$) &  $0.239_{-0.129}^{+0.219}$ & $0.503_{-0.348}^{+0.308}$ & $-0.046_{-0.135}^{+0.181}$ & 1.17 & 52.75 & 44.97 &$0.934_{-0.133}^{+0.178}$  \\
\hline
flat $\Lambda$CDM+$z=z_{\rm obs}$(SN+BAO--Ly$\alpha$) & $0.293_{-0.030}^{+0.034}$ & $\Omega_{\Lambda}=1-\Omega_{\rm m}$ & $\alpha=0$ & 1.12 & 46.37 & 41.70 & $0.985_{-0.024}^{+0.025}$\\
 & & & & & & & \\
flat $\Lambda$CDM+$z\neq z_{\rm obs}$(SN+BAO--Ly$\alpha$) & $0.192_{-0.094}^{+0.296}$ & $\Omega_{\Lambda}=1-\Omega_{\rm m}$ & $0.113_{-0.246}^{+0.174}$ & 1.14 & 49.56 & 43.34 & $1.093_{-0.238}^{+0.167}$ \\
\hline
oCDM+$z=z_{\rm obs}$(SN+BAO--Ly$\alpha$) & $0.000_{-0.000}^{+0.023}$ & $\Omega_{\Lambda}=0$ & $\alpha=0$ & 1.67 & 64.04 & 59.37 & $0.902_{-0.019}^{+0.016}$ \\
 & & & & & & & \\
oCDM+$z\neq z_{\rm obs}$(SN+BAO--Ly$\alpha$) & $0.369_{-0.190}^{+0.260}$ & $\Omega_{\Lambda}=0$ & $-0.205_{-0.054}^{+0.049}$ &1.13 & 49.37 & 43.15 &$0.773_{-0.047}^{+0.042}$\\
\hline
flat CDM+$z=z_{\rm obs}$(SN+BAO--Ly$\alpha$) & $\Omega_{\rm m}=1$ & $\Omega_{\Lambda}=0$ & $\alpha=0$ & 8.23 & 278.8 & 275.7 & $0.777_{-0.014}^{+0.014}$ \\
 & & & & & & & \\
flat CDM+$z\neq z_{\rm obs}$(SN+BAO--Ly$\alpha$) & $\Omega_{\rm m}=1$ & $\Omega_{\Lambda}=0$ & $-0.310_{-0.016}^{+0.017}$ &1.25 & 50.57 & 45.90 &$0.681_{-0.014}^{+0.015}$\\
\end{tabular}
\caption{Constraints on the parameters of cosmological models with the standard 
redshift mapping ($z=z_{\rm obs}$) or with redshift remapping given by eq.~(\ref{remapping}) ($z\neq z_{\rm obs}$) and 
parametrized by $\alpha$, based on a joint analysis of the SN and BAO data without Ly$\alpha$ forest measurements at $z=2.34,\;2.36$. 
The table contains best-fitting values and $1\sigma$ confidence intervals of the parameters (density parameters $\Omega_{\rm m}$ and $\Omega_{\Lambda}$, 
parameter $\alpha$ of the redshift remapping and the normalization of the BAO signal, $M_{\rm BAO}$), reduced $\chi^{2}$ ($\chi^{2}/{\rm dof}$), the Bayesian information criterion 
(BIC) and the Akaike information criterion (AIC). The models comprise an open $\Lambda$CDM model 
(o$\Lambda$CDM, see also Fig.~\ref{openLCDMa}), a flat $\Lambda$CDM model (flat $\Lambda$CDM, $\Omega_{\rm m}+\Omega_{\Lambda}=1$) 
an open CDM model without dark energy (oCDM, $\Omega_{\Lambda}=0$) and the Einstein--de Sitter model (flat CDM, $\Omega_{\rm m}=1$ 
and $\Omega_{\Lambda}=0$).
}
\label{fitsa}
\end{center}
\end{table*}

Table~\ref{fits} summarizes all fits in terms of the best-fitting values of the parameters and the corresponding $1\sigma$ confidence intervals, goodness of fit and the two information criteria described above. The two top rows of the table present results obtained for an open 
$\Lambda$CDM model (o$\Lambda$CDM, see also Fig.~\ref{openLCDM}), the middle two rows for a flat $\Lambda$CDM model (flat $\Lambda$CDM, $\Omega_{\rm m}+\Omega_{\Lambda}=1$) and the two bottom rows for an open CDM model without dark energy (oCDM, $\Omega_{\Lambda}=0$). In every case, we consider redshift remapping 
parametrization given by eq.~(\ref{remapping}) or the standard redshift mapping with $z_{\rm obs}=z$.

The BAO measurements based on the Ly$\alpha$ forest of BOSS quasars were found to be in tension with the Planck cosmology \citep{Del2015}. Since the nature of this tension is not clear, it is instructive to scrutinize the impact of these measurements on our 
fits. Fig.~\ref{openLCDMa} shows constraints on the parameters of an open $\Lambda$CDM model from a joint analysis of the SN and BAO data excluding the Ly$\alpha$ forest measurements (distances and Hubble parameters at $z_{\rm obs}=2.34,\;2.36$). Table~\ref{fitsa} contains results of fitting models with different priors on dark energy and curvature, analogous to Table~\ref{fits}. Here we also consider the Einstein--de Sitter model.

\begin{figure*}
\begin{center}
    \leavevmode
    \epsfxsize=16cm
    \epsfbox[85 70 860 530]{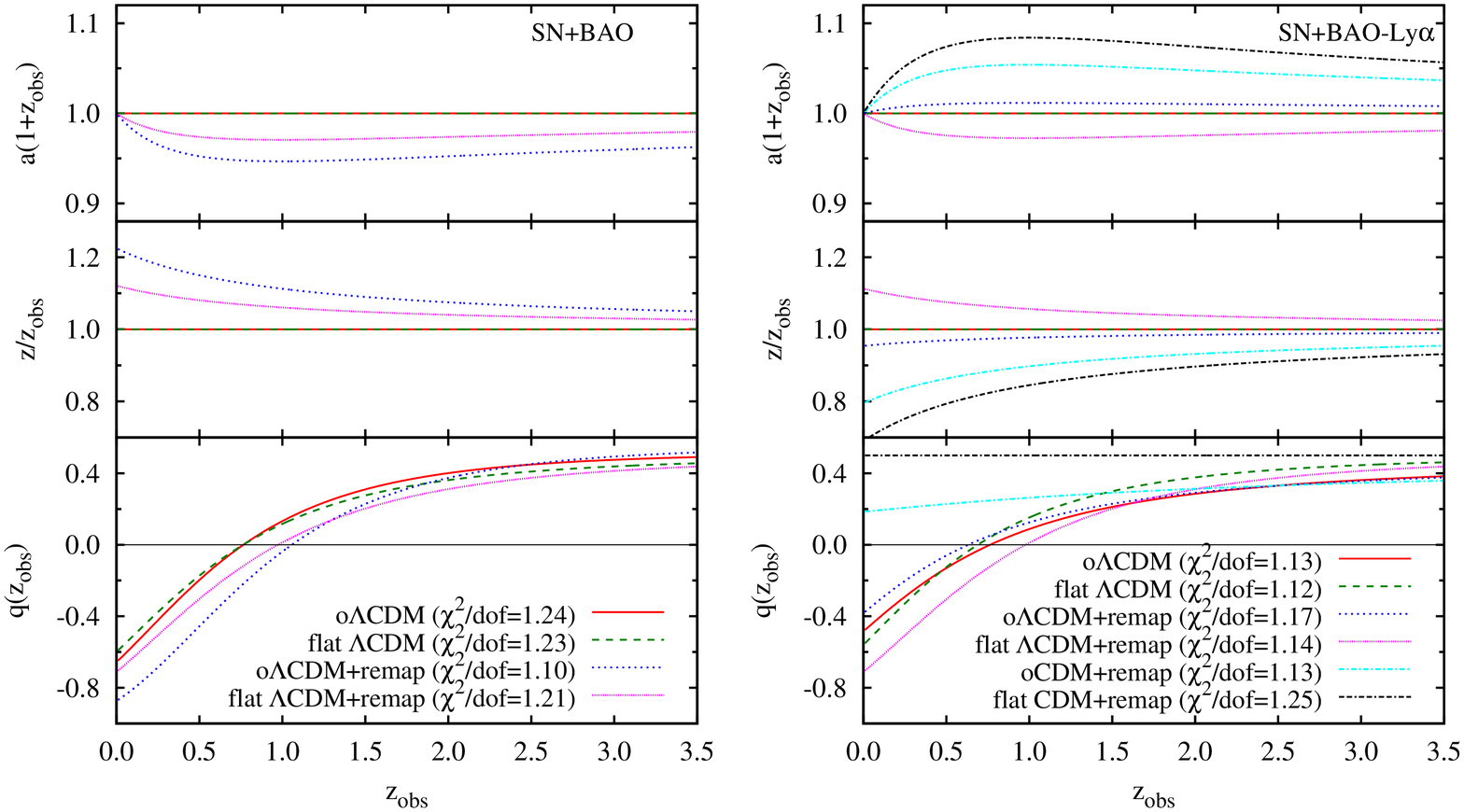}
\end{center}
\caption{Mapping between cosmic scale factor $a$ (or the model redshift $z=1/a-1$) and the observed redshift $z_{\rm obs}$, and the corresponding deceleration parameter in cosmological models with $\chi^{2}/{\rm dof}\leq1.25$. The lines show redshift profiles obtained for best-fitting parameters. The left-hand panels show the results for models fitted to the SN and BAO data sets (see Table~\ref{fits}), while the right-hand panels display models fitted to the SN and BAO data excluding the BAO Ly$\alpha$ forest measurements of BOSS quasars (see Table~\ref{fits}). Redshift remapping with $\alpha<0$ ($z<z_{\rm obs}$) can effectively replace dark energy and yield a decelerating expansion at the present time.
}
\label{mapping}
\end{figure*}

The obtained constraints on parameter $\alpha$ based on combining BAO and SN data do not reveal statistically significant deviation from the standard redshift mapping. The overall improvement of a cosmological fit with an open $\Lambda$CDM model is marginal and the current data do not seem to favour the new redshift remapping model. Both Fig.~\ref{openLCDM} and Fig.~\ref{openLCDMa} clearly demonstrate that redshift remapping is strongly degenerated with the dark energy density. Models with low dark energy density or no dark energy at all require $\alpha<0$ and thus $z=1-1/a>z_{\rm obs}$, while those with high dark energy density $a>0$ and thus $z=1-1/a>z_{\rm obs}$. Yet, redshift remapping alleviates a well visible discrepancy between the best-fitting $\Lambda$CDM model with the standard redshift mapping and the BAO measurements from the Ly$\alpha$ forest of BOSS quasars at $z_{\rm obs}=2.34$ and cross-correlation with quasars at $z_{\rm obs}=2.36$ (see Fig.~\ref{residuals}). This discrepancy is a manifestation of already reported tension between the Ly$\alpha$ forest BAO measurements and the Planck cosmology, with significance estimated at $2.5\sigma$ \citep{Del2015,Car2015}. Comparing fits based on the combined SN and BAO data including or excluding the Ly$\alpha$ forest data points (compare results in Table~\ref{fits} and Table~\ref{fits}) makes evident that it is the Ly$\alpha$-forest BAO measurement what gives rise to a positive value of the parameter $\alpha$ at a $2\sigma$ confidence level. The weak statistical significance of this result does not allow to judge if this is a real signature of a non-trivial redshift remapping. Undoubtedly, future measurements of BAO in an optimal redshift range $1<z_{\rm obs}<2$ will shed light on this problem.

Since our analysis does not signify statistically significant deviation from the standard redshift mapping, it is not surprising that the constraints on cosmological parameters are consistent with those obtained assuming the standard mapping (see Table~\ref{fits} and Table~\ref{fits}). However, the new redshift remapping changes substantially degeneracies between them. As shown in the left-hand panels of Fig.~\ref{openLCDM} and Fig.~\ref{openLCDMa}, marginalization over parameter $\alpha$ expands the probability contours on the $\Omega_{\rm m}-\Omega_{\Lambda}$ plane nearly orthogonally to the degeneracy axis emerging from fitting a $\Lambda$CDM model with the standard redshift mapping (see the inset panels). This effectively rotates the axis of degeneracy between $\Omega_{\rm m}$ and $\Omega_{\Lambda}$ from a direction nearly perpendicular to a line of flat models towards a direction of flat models (more vertical for constraints based on the SN data set and more horizontal for the BAO data set). The apparent reorientation of the degeneracy axis shows how large is the impact of the standard redshift mapping on the determination of cosmological parameters. Our results demonstrate that relaxing the assumption of the standard relation between observed redshift $z_{\rm obs}$ and cosmic scale factor $a$ makes probability contours on the $\Omega_{\rm m}-\Omega_{\Lambda}$ plane more aligned with those from an analysis of the CMB observations \citep[see e.g.][]{Pla2014}.

Comparing results from Fig.~\ref{openLCDM} and Fig.~\ref{openLCDMa} leads to the conclusion that current SN and BAO data, when analysed in a framework of a cosmological model with redshift remapping, do not constrain dark energy unless the BAO 
data include the current measurements from the Ly$\alpha$ forest of BOSS quasars. This is a consequence of a strong degeneracy between redshift remapping and dark energy density,  as shown in right-hand panels of both figures. Limiting our consideration to cosmological fits 
based on SN and BAO data without the Ly$\alpha$ forest data points (see Fig.~\ref{openLCDMa} and Table~\ref{fitsa}), we find that open and flat CDM models $-$ without dark energy 
$-$ assuming redshift remapping provides remarkably as good fits 
to the SN and BAO data as a $\Lambda$CDM model with standard mapping. Fig.~\ref{residuals} shows directly that residuals for an open CDM model with redshift remapping are indistinguishable from a $\Lambda$CDM model with the standard mapping ($z_{\rm obs}=z$). Just from a statistical point of view, this means that redshift remapping can in general mimic kinematical effects of space expansion driven by dark energy. Fitting a CDM model to observationally determined distances and Hubble parameters requires involving redshift remapping with $\alpha<0$ and thus $z=1-1/a>z_{\rm obs}$. The flat model (the Einstein--de Sitter model) needs slightly smaller $z$-to-$z_{\rm obs}$ ratio than an open one. The matter density parameter in an open CDM model is fully consistent with constraints based on observations of galaxy clusters \citep{Bah1998,Roz2010,All2011}

Fig.~\ref{mapping} shows mapping between observed redshifts $z_{\rm obs}$ and cosmic scale factor, as constrained by both SN and BAO data. The profiles are calculated for best-fitting parameters of models with reduced $\chi^{2}$ no larger than $1.25$. Although the shape of the mapping depends partly on the adopted parametrization of our model, its normalization is directly constrained by the data and therefore we think it captures a generic feature expected for a family of redshift remappings with monotonic functions $f(z_{\rm obs})$. The apparent trends with $z>z_{\rm obs}$ ($z<z_{\rm obs}$) implies that the standard redshift mapping \textit{overestimates} (\textit{underestimates}) the scale factor corresponding to observed redshifts $z_{\rm obs}$. The right-hand panels demonstrate that emergence 
of dark energy in cosmological fits may result from a necessity of compensating underestimation of the scale factor by the standard redshift mapping.

The bottom panels of Fig.~\ref{mapping} show how the new redshift remapping modifies the deceleration parameter. The most striking effect is visible for models in which redshift remapping allows for a significance change of the density parameters, e.g. flat and open CDM models. A more subtle effect is a shift of the observed redshift corresponding to the moment of a vanishing deceleration (for $\Lambda$CDM models).

\section{Discussion}

\begin{figure}
\begin{center}
    \leavevmode
    \epsfxsize=8cm
    \epsfbox[72 65 455 460]{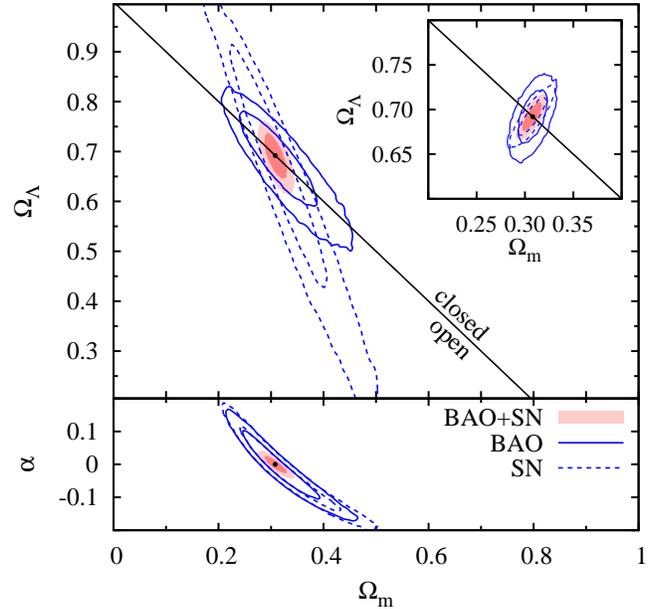}
\end{center}
\caption{Expected observational constraints on the parameters of open $\Lambda$CDM models with redshift remapping given by eq.~(\ref{remapping}), based on the BAO forecast observations by DESI and SN Ia by LSST, and generated for a fiducial $\Lambda$CDM model (marked with the black point) with the standard remapping, i.e. $z_{\rm obs}=z$. The solid blue and dashed blue lines show results from BAO and SN, the shaded contours 
from analysis combining both data sets ($1\sigma$ and $2\sigma$ contours). The inset panel shows constraints on the density parameters for an open $\Lambda$CDM model assuming the standard redshift mapping ($z_{\rm obs}=z$). Future constraints will retain nearly the same degeneracy axes as those shown in Fig.~\ref{openLCDM}. The expected precision in the measurement of $\alpha$ parameter will reach $0.02$ (fivefold improvement of the current precision).
}
\label{forecast}
\end{figure}

Future measurements of BAO and observations of Type Ia supernovae will enable us to test the standard redshift mapping with a much higher precision. To assess the constraining power of upcoming surveys, we analyse mock data generated for the expected measurements of the BAO signal by the Dark Energy Spectroscopic Instrument\footnote{http://desi.lbl.gov} (DESI) and for the anticipated number of Type Ia supernovae to be discovered by the Large Synoptic Survey Telescope\footnote{http://www.lsst.org} (LSST). DESI will measure the BAO distance scale in a large redshift range, from almost $z_{\rm obs}=0$ up to redshift 3.5, using galaxy and quasar spectra over an area of 14,000 deg$^{2}$ \citep[see][for the DESI BAO data forecast]{BigBOSS,DESI,Font-Ribera14}. For the SN data forecast 
we assume that LSST will observe $\sim 10^{5}$ supernovae \citep{LSST2009}. Considering the best case scenario, we neglect all possible systematic errors and obtain the data forecast by means of rescaling the current covariance matrix according to a ratio of the current to future number of supernovae. Fig.~\ref{forecast} shows the anticipated constraints on the model parameters calculated for mock data generated for a fiducial $\Lambda$CDM model with the standard redshift mapping. Unsurprisingly, future constraints will retain nearly the same degeneracy axes as those shown in Fig.~\ref{openLCDMa}. The expected precision in the measurement of parameter $\alpha$ will reach $0.02$, which yields a fivefold improvement of the current constraints.

Since redshift remapping modifies angular diameter distances, we expect that additional constraints can be obtained from observations of strong lensing images of background galaxies produced by massive galaxy clusters. This technique has been already proved to be a complementary cosmological probe capable of measuring dark energy equation of state \citep{Jul2010} or testing alternative models to the late time cosmic acceleration of the Universe \citep{Mag2015}.

Placing constraints on redshift remapping cannot be disjoined from measuring cosmological parameters such density parameters. Our results clearly show that redshift remapping affect constraints on cosmological parameters and is even capable of eliminating the dark energy component in a given cosmological model. This in turn may modify 
predictions for the CMB anisotropies. Therefore, we conclude that future tests of the standard redshift mapping should not be limited to low-redshift cosmological probes such as SN and BAO, but they should also include CMB observations. This approach will not only ensure consistent analysis of all relevant cosmological probes, but it will also allow us to extend the test to a redshift of the last scattering surface.

The model of redshift remapping employed in this work is not derived from first principles. We also assume that the Friedmann equation remains unchanged regardless of deviation from the standard redshift mapping. This might not be fully justified for some theoretical models predicting $z_{\rm obs}\neq z$. Therefore, further advances in testing and understanding the redshift$-$scale factor mapping relation should also invoke theoretical models predicting specific forms of redshift remapping. It will be also instructive to interpret our results in a framework of general-relativistic models with scalar averaging \citep[backreaction models; see e.g.][]{Lar2009,Rou2013,Lav2013}.

\section{Summary}
We present the first self-consistent observational test of mapping between observed cosmological redshifts and cosmic scale factor (redshift remapping). The test relies on placing simultaneous observational constraints on standard cosmological parameters and a phenomenological model generalizing the standard relation between cosmic scale factor and observed redshifts, as given by eq.~(\ref{remapping}). Using current observations of Type Ia SN and BAO, we measure a deviation from the standard redshift mapping and its impact on 
constraining the dark energy density parameter. Our main results can be summarized as follows.
\begin{description}
\item[i.] Joint analysis of SN and BAO data in an open $\Lambda$CDM model does not signify statistically significant deviation from the standard redshift mapping ($z_{\rm obs}=z$). 
Combined constraints from SN and BAO data sets yield $\alpha=0.225_{-0.120}^{+0.083}$, which points to a weak ($\sim2\sigma$ confidence) trend of $z=1/a-1>z_{\rm obs}$. 
This result turns out be heavily dependent on the Ly$\alpha$ forest BAO measurements. When excluding these BAO data points, constraints on $\alpha$ change into $\alpha=-0.046_{-0.135}^{+0.181}$, what points to a weak indication of a reversed trend with $z=1/a-1<z_{\rm obs}$.
\item[ii.] Redshift remapping alleviates the reported $2.5\sigma$ tension between the standard $\Lambda$CDM and the Ly$\alpha$ forest BAO measurements. 
The fit requires a positive value of $\alpha$ and thus $z=1/a-1>z_{\rm obs}$.
\item[iii.] Relaxing the assumption of standard redshift mapping in cosmological fits changes substantially the degeneracy between cosmic density parameters. Marginalization over redshift remapping effectively rotates the axis of degeneracy on the $\Omega_{\rm m}-\Omega_{\Lambda}$ plane towards a line of flat models (BAO data) or 
towards a vertical line $\Omega_{\rm m}\approx 0.2$ (SN data and BAO data without the Ly$\alpha$-forest measurements).
\item[iv.] Redshift remapping is strongly degenerated with dark energy density parameter. Low-density dark energy models, or no dark energy at all, require $\alpha<0$ and thus $z=1/a-1>z_{\rm obs}$.
\item[v.] The proposed redshift remapping parametrization can effectively replace dark energy. SN and BAO data analysed in a framework of a cosmological model with redshift 
remapping do not constrain dark energy unless the BAO data include the Ly$\alpha$ forest BAO measurements. When excluding these BAO Ly$\alpha$ forest observations, an open CDM model with redshift remapping is able to fit the SN and BAO data equally well as a $\Lambda$CDM model with the standard redshift mapping ($z_{\rm obs}=z$). The resulting redshift remapping is characterized by $\alpha=-0.205_{-0.054}^{+0.049}$ and thus $z=1/a-1>z_{\rm obs}$. Comparable goodness of fit is achieved for the Einstein--de Sitter model.
\end{description}

Finally, we highlight and conclude in this work how the future generation of BAO and SN experiments, such as DESI, LSST, \textit{4MOST}\footnote{https://www.4most.eu} and 
\textit{Euclid}\footnote{http://sci.esa.int/euclid/} will enable us to test cosmology models and the redshift$-$scale factor relation with high accuracy/precision consistency. Analysis and prospects of forecast observations of both surveys suggest that redshift remapping will be constrained to within a 2 per cent precision.

\section*{Acknowledgements}
We would like to thank Tom Abel, Jorge Alfaro, Maciej Bilicki, Roger Blandford, Tamara Davis, Daniel Eisenstein, Bernard Jones, Anatoly Klypin, Claudio Llinares, Boud Roukema, Alberto Salvio, Miguel Zumalac\'{a}rregui for useful discussions; and Franco Albareti, David Bacon, Chris Blake, Johan Comparat, Sebastian Szybka, Marek Szyd{\l}owski  for critical reading of the manuscript. Our gratitude to Anatoly Klypin for suggesting the generalized $\alpha$-parametrization form of the redshift$-$scale factor remapping relation used in this work. RW acknowledges support through the Porat Postdoctoral Fellowship. The Dark Cosmology Centre is funded by the Danish National Research Foundation. FP acknowledges support from the Spanish MICINNs Consolider-Ingenio 2010 Programme under grant MultiDark CSD2009-00064, MINECO Centro de Excelencia Severo Ochoa Programme under grant SEV-2012-0249, and grant AYA2014-60641-C2-1-P.  FP wishes to thank KIPAC for the hospitality during the development of this work.

\bibliography{master}

\end{document}